\documentclass[aps,preprint,showpacs,superscriptaddress,groupedaddress]
{elsarticle}
\usepackage{bm}       
\usepackage{amssymb}
\usepackage{amsmath}     

\usepackage{hyperref} 
\usepackage{psfrag}
\usepackage[cp1250]{inputenc}

\usepackage{graphicx}

\DeclareMathAlphabet{\mathpzc}{OT1}{pzc}{m}{it}

\def\thebiblio#1{
\begin{center}\bf \large References
\end{center}
\list
{[\arabic{enumi}]}{\settowidth\labelwidth{#1.}\leftmargin\labelwidth
 \advance\leftmargin\labelsep
 \usecounter{enumi}}
 \def\newblock{\hskip .11em plus .33em minus -.07em}
 \sloppy
 \sfcode`\.=1000\relax}

\newcommand{\lsim}{\mbox{\raisebox{-.9ex}{~$\stackrel{\mbox{$<$}}{\sim}$~}}}
\newcommand{\gsim}{\mbox{\raisebox{-.9ex}{~$\stackrel{\mbox{$>$}}{\sim}$~}}}

\numberwithin{equation}{section}

\begin{document}

\title{Initial conditions for inflation}

\author[lancaster]{Konstantinos Dimopoulos} 
\author[krakow]{and Micha{\l} Artymowski}

\address[lancaster]
{Consortium for Fundamental Physics, Physics Department, 
Lancaster University, Lancaster LA1 4YB, UK\\}

\address[krakow]
{Jagiellonian University,  
{\L}ojasiewicza 11, 30-348 Krak{\'o}w, Poland\\}

\begin{abstract}
Within the $\alpha$-attractors framework we investigate scalar 
potentials with the same pole as the one featured in the kinetic term. We show 
that, in field space, this leads to directions without a plateau. Using this, 
we present a proposal, which manages to overcome 
the initial conditions problem of inflation with a plateau. An earlier period of 
proto-inflation, beginning at Planck scale, accounts for the Universe
expansion and arranges the required initial conditions for inflation on the 
plateau to commence. We show that, if proto-inflation is power-law, it does not
suffer from a sub-Planckian eternal inflationary stage, which would otherwise 
be a problem. A simple model realisation is constructed in the context of 
$\alpha$-attractors, which can both generate the inflationary plateau and the 
exponential slopes around it, necessary for the two inflation stages. Our 
mechanism allows to assume chaotic initial conditions at the Planck scale for 
proto-inflation, it is generic and it is shown to work without fine-tuning.
\end{abstract}

\date{\today}

\maketitle

\tableofcontents

\section{Introduction} \label{sec:Introduction}

Cosmic inflation \cite{Guth:1980zm,Starobinsky:1980te} accounts for the 
fine-tuning problems of hot big bang cosmology (horizon and flatness), 
explains the origin of structure (primordial curvature perturbation) and it is 
fully consistent with observational data \cite{Ade:2015lrj}. Recent observations
are setting stronger bounds on tensor-to-scalar ratio $r$, limiting the maximal
scale of inflation and favour flatter inflationary potentials, featuring an 
inflationary plateau, which may be motivated by modifications of gravity 
\cite{Starobinsky:1980te}, quantum field theory \cite{Tmodel,Emodel}
or both \cite{Bezrukov:2007ep}. 

Inflation models with a plateau are characterised by a scalar potential limited
from above by the energy scale of inflation, which due to the constrains from 
observations cannot be bigger than the scale of a grand unified theory (GUT)
\mbox{$m_{_{\rm GUT}}\sim 10^{16}\,$GeV}. This gives the rise to the problem of 
initial conditions, described in e.g. Ref.~\cite{Ijjas:2013vea}. The issue is 
the following: Let us assume that the Planck scale $M_P\sim 10^{19}\,$GeV is a 
natural scale for setting initial conditions for the pre-inflationary Universe. 
{Assuming an expanding Universe,}
since the density scale of the inflationary plateau is at least 
$(\frac{M_P}{m_{_{\rm GUT}}})^4\sim 10^{12}$ times smaller, there may be a long 
period of a decelerated expansion of the pre-inflationary Universe, during 
which the cosmological horizon is dominated by inhomogeneities. To avoid this, 
one needs to assume that the Universe is homogeneous over exponentially many 
horizons at the Planck scale, which leads to massive fine-tuning of initial 
conditions. {This is because for a long time it has been considered that 
inflation needs a homogeneous patch of at least a Hubble volume to begin with.}

However, recently, it has been argued that, even inhomogeneous (and 
anisotropic) initial conditions for the large field inflation may still allow 
inflation to take hold \cite{East:2015ggf,Clough:2016ymm,Kleban:2016sqm}.
Nevertheless, the analysed scenarios assume an initially non-contracting space,
which in the case of exponentially many causally disconnected regions seems 
to be another source of massive fine-tuning.\footnote{In fact, a varying 
extrinsic curvature (which determines whether space is expanding or contracting)
has been partially considered only in Ref.~\cite{Clough:2016ymm} and only 
under special initial conditions 
(e.g. constant initial velocity of the inflaton). Space was assumed to be
predominantly expanding, while the authors acknowledge that
``In general, we should expect the local expansion rate to be a function of 
spatial position that can be both initially expanding or collapsing. \ldots 
We reserve the general case for future work.''}

In fact, the problem can be more acute because the Planck-scale Universe lacks 
the initial boost towards the expansion of space, 
so an initially expanding space without
inflation is non-trivial to postulate. Refs.~\cite{East:2015ggf,Clough:2016ymm} 
considering inhomogeneous initial conditions, assume expansion (at least on 
average) such that they stay clear from the quantum complications of spacetime 
foam at the Planck scale. However, one can argue that, because in spacetime 
foam the extrinsic curvature can change by quantum fluctuations, one could not 
have sustained expansion, so no net expansion, i.e. on average $\dot a=0$. Now, 
to get from no expansion to expansion you need $\ddot a>0$, that is inflation.
Thus, it is inflation that produces the Universe expansion. Without it, the 
Universe remains at the Planck scale. 

Indeed, this is supported by some early work of Brout, Englert and Gunzug \cite{Brout:1979bd}, and Zel'dovich and Vilenkin
\cite{Zelkin} (see also Ref.~\cite{mukhanov}). More recently, 
results from the theory of Causal Dynamical Triangulations 
\cite{Ambjorn:2012jv} also indicate the crucial role of high energy 
cosmological constant (which in the realistic case should be replaced by the 
flat inflationary potential) in the process of creating a classical Universe 
from the quantum foam. Therefore, we see that we need to start inflation at the
Planck scale at density much larger than the inflationary plateau.

To solve or ameliorate the problem of initial conditions one can include the 
internal curvature of the FRW metric \cite{Guth:2013sya,Dalianis:2015fpa}, 
assume compact topology \cite{Linde:2004nz}, consider the Jordan frame Planck 
scale as a physical one \cite{Gorbunov:2014ewa} (for modified gravity inflation)
or include a proto-inflationary phase, which would homogenise the Universe at 
the Planck scale \cite{Hamada:2014wna,Carrasco:2015rva,Artymowski:2016ikw}.

In the latter case case of multi-phase inflation, the scenario suffers from the
problem of extensive eternal proto-inflation; when quantum corrections 
overwhelm the classical evolution of the proto-inflaton field and, in
many horizons, the inflaton cannot reach its minimum. This issue is especially 
dangerous when the degree of freedom that is responsible for proto-inflation is 
eternally inflating at sub-Planckian density. While undergoing eternal 
inflation, the proto-inflaton cannot stop inflating, while the plateau inflaton
may keep rolling towards its minimum, leaving no space for a GUT-scale 
accelerated expansion. 

In this paper we propose another idea, namely a two field inflationary 
scenario with an initial power-law inflation \footnote{The model of power-law 
inflation has firs been studied in Ref. \cite{Lucchin:1984yf}} and a subsequent plateau 
inflation as proto-inflation and GUT-scale plateau-inflation respectively. We 
show that, even though starting at the Planck scale, power-law proto-inflation 
evades the sub-Planckian eternal inflation problem and lasts only a limited 
number of e-folds, such that the system safely lands on the plateau of the 
slow-roll GUT-scale plateau-inflation. 
A similar proto-inflation model is presented in Ref.~\cite{Carrasco:2015rva}.
In contrast to that proposal, our model is generic, in that any plateau model 
of inflation can be accommodated, while our proto-inflaton and plateau-inflaton
fields are unrelated and may correspond to degrees of freedom of different 
sectors of the theory. Our proposal can be naturally realised in the 
context of the $\alpha$-attractors \cite{Tmodel,Emodel}, in a different way
than in Ref.~\cite{Carrasco:2015rva}. 

We consider natural units, where $c=\hbar=1$ and Newton's gravitational 
constant is \mbox{$8\pi G=m_P^{-2}$}, with 
\mbox{$m_P\equiv M_P/\sqrt{8\pi}=2.43\times 10^{18}\,$GeV} 
being the reduced Planck mass.


\section{The model} \label{sec:Model}

Our proposal is that the inflationary scalar potential is of the form
\begin{equation}
V=V(\varphi)+V(\psi)\,,
\end{equation}
where $V(\varphi)$ is featuring the inflationary plateau with 
\mbox{$V(\varphi)\lsim m_{_{\rm GUT}}^4$} and is responsible for potentially long 
slow-roll, GUT-scale plateau-inflation, which generates the observed curvature 
perturbation, while $V(\psi)$ is responsible for the limited 
proto-inflation period which accounts only for the initial conditions and is 
negligible afterwards. Initially, we expect
\begin{equation}
V(\varphi)\lsim m_{_{\rm GUT}}^4\ll V(\psi)\lsim m_P^4 \, .
\end{equation}
We argue that proto-inflation can be power-law with the scale factor growing as 
$a\propto t^p$, where $p$ is a constant parameter (with $p>1$ for inflation). 
Thus, during this period, 
\mbox{$V(\psi)\propto\exp\left(\sqrt{\frac{2}{p}}\psi/m_P\right)$}, where 
without loss of generality we have chosen $\psi>0$ \cite{DavidBook1}. 
In that way, proto-inflation may last only a limited number of e-folds, while 
sub-Planckian eternal inflation can be altogether avoided.

\subsection{%
The danger of an extensive diffusion zone on the hill}

We motivate considering power-law inflation by avoidance of an extensive 
diffusion zone, i.e. one which corresponds to sub-Planckian energies, on the 
hills of the plateau valley. 

The expectation value of a scalar field $\phi$ during inflation changes due to 
its classical slow-roll evolution as\footnote{The prime and the dot denote 
derivative with respect to the inflaton field and the cosmic time respectively.}
\mbox{$|\dot\phi|\simeq|V'|/3H$} and due to its quantum fluctuations by
\mbox{$\delta\phi/\delta t=H^2/2\pi$}, i.e. given by the Hawking temperature per
Hubble time. Many inflation models have regions (called diffusion zones) where
the quantum fluctuations overwhelm the classical evolution of the field(s). 
Inside the diffusion zone the scalar field is oblivious of the potential so it 
does not roll towards its minimum, while it may exit the diffusion zone only via
chaotic quantum fluctuations. If, during inflation, the system finds itself 
inside a diffusion zone, then it undergoes eternal inflation \cite{eternal}. 
This means that, even though the system may typically exit the diffusion zone 
eventually, there will always be locations in physical space where the system 
remains trapped in the diffusion zone so that inflation continues.

In our setup we consider two stages on inflation. The proto-inflaton field 
$\psi$ is driving the initial stage, taking the system from Planckian  
density down to GUT-scale density, where the second stage of inflation 
takes place, driven by the plateau-inflation field $\varphi$. Thus, the 
GUT-scale plateau direction (parametrised by $\varphi$) corresponds to a valley
in field space, while its walls (the hills) correspond to the orthogonal 
proto-inflaton direction (parametrised by $\psi$), see Fig.~\ref{fig}.

The problem with an extensive diffusion zone on the hill is the following.
During proto-inflation, the plateau-inflaton field $\varphi$, being light (the 
plateau is very flat), undergoes intense quantum fluctuations that send it to 
large values along the plateau. If proto-inflation is eternal then the build-up
of the $\varphi$ condensate becomes very large and so the typical expectation 
value of $\varphi$ moves further down the plateau valley and away from minimum 
(taken at $\varphi=0$). However, the plateau-inflaton is not light near the 
minimum, where the valley becomes steep and curved. When the plateau-inflaton 
finds itself in this region it no more undergoes eternal inflation but instead 
it does roll towards its minimum.\footnote{{Indeed, in 
Refs.~\cite{East:2015ggf,Clough:2016ymm} it is shown that, if the plateau 
inflaton, in a given region, finds itself in the potential minimum, this
drags the field into the minimum in neighbouring areas too.}}
Now, this region is larger when $H$ is 
smaller, because the slope of the potential along the valley can overcome 
the ``quantum kicks''. Thus, if the diffusion zone on the hill is extended,
then eternal inflation on the hill may occur even for $H$ much smaller than 
$m_P$, in which case the region of where the plateau-inflaton allows slow-roll 
towards the minimum (when $V'(\varphi)>H^3$) is enlarged. 

This means that there is a preference against an extended diffusion zone on the
hill, because, were there one, then the proto-inflaton $\psi$ could still be in 
it (so undergoing eternal inflation) for sub-Planckian density. However, the 
more $H$ decreases the more the diffusion zone in the valley for the 
plateau-inflaton $\varphi$ withdraws from the minimum, so the more likely it 
becomes that $\varphi$ finds itself outside its diffusion zone and begins to 
roll towards the minimum, while $\psi$ is still eternally inflating on 
the hill. In this case, there may be no plateau-inflation left, once the 
proto-inflation finishes. 

As a simple choice, consider a proto-inflaton field with 
$V(\psi)=\frac12m^2\psi^2$ potential such that the hills of the valley 
correspond to the quadratic rise of the potential. The eternal inflation regime 
corresponds to $\psi\gtrsim m_P\sqrt{m_P/m}$, which is well below the Planck 
density scale, for which $\psi\sim m_P^2/m$. Taking $m\sim m_P$ would push the 
eternal inflation limit towards the Planck scale but the potential would not be
able to generate many inflationary e-folds, since quadratic chaotic inflation 
ends when $\psi\sim m_P$ regardless of~$m$. Therefore, this choice of $V(\psi)$
results to an extended diffusion zone, which is undesirable. 

In contrast, in the case of power-law inflation it can be shown that
\mbox{$|\dot\psi|=\sqrt{\frac{2}{p}}m_PH$}, which should be contrasted
with \mbox{$(\delta\psi/\delta t)=H^2/2\pi$}. Thus, to be outside
the diffusion zone we need \mbox{$\rho=3H^2m_P^2<(\frac{24}{p})\pi^2 m_P^4$}. 
This is comparable to the Planck density \mbox{$M_P^4=(8\pi)^2m_P^4$} if 
$p$ is not too large. Therefore, in power-law proto-inflation, the diffusion 
zone is confined to the Planck scale. So, after leaving the Planck scale, 
proto-inflation can avoid eternal inflation.

\subsection{ 
Multifield inflation from $\alpha$-attractors}

In supergravity, the general expectation is that scalar fields have 
non-canonical kinetic terms. This is because the kinetic terms are determined by
the K\"{a}hler metric, which can be complicated when the K\"{a}hler potential is
not minimal. Thus, inflation model-building in supergravity and string theory
features frequently non-canonical kinetic terms for the scalar fields, such as 
the dilaton or the complex structure moduli fields.

In recent years, it was shown \cite{Tmodel,Emodel} that a non-canonical kinetic 
term with a pole may be responsible for generating a plateau in the 
inflationary potential. This effect, corresponding to families of inflation 
models known as $\alpha$-attractors, ``stretches'' the potential around the 
pole of the kinetic term \cite{Linde:2016uec}. 
However, it can be shown that a pole of the kinetic term does not necessarily 
lead to a plateau in the scalar potential, when the latter also features 
{\em the same pole}. We exploit this possibility when generating the potential 
for our power-law proto-inflation. {In fact the possibility of obtaining 
inflationary potentials without a plateau in the $\alpha$-attractors scheme is 
a secondary motivation for our paper. In what follows, we show how different 
inflationary scenarios can emerge from poles of scalar potential.}

Let us consider a multifield scenario in which the scalar potential of the
proto-inflaton field features the same pole as its kinetic term, namely
\begin{equation}
\mathcal{L}=
\frac{1}{2}R + 
\frac{\frac{1}{2}(\partial \phi)^2}{\left(1-\frac{\phi^2}{6\alpha}\right)^2} + 
\frac{\frac{1}{2}(\partial s)^2}{\left(1-\frac{s^2}{6\beta}\right)^2} - 
\frac{\frac{1}{2}m_\phi^2 \phi^2}{\Big(1+\frac{\phi}{\sqrt{6\alpha}}\Big)^2} -
\frac{\frac{1}{2}m^2 s^2}{1-\mbox{\Large $\frac{s^2}{6\beta}$}} \,, 
\label{eq:alphaattractor}
\end{equation}
where $\alpha$ and $\beta$ are positive constants and we have set $m_P=1$. 
Note that the fields need not be directly coupled. The following field 
redefinitions
\begin{equation}
\phi = \sqrt{6\alpha}\tanh\frac{\varphi}{\sqrt{6\alpha}} 
\quad{\rm and}\quad
s = \sqrt{6\beta}\tanh\frac{\psi}{\sqrt{6\beta}}\,,
\label{eq:alphakin}
\end{equation} 
lead to canonical kinetic terms. The potential as a function of $\varphi$ 
stretches near the pole in \eqref{eq:alphaattractor}, which gives a rise to 
flatness as the pole is transposed to infinity. For the canonical fields, 
we have
\begin{equation}
V(\varphi) = V_C \left(1-e^{-\sqrt{\frac{2}{3\alpha}}\varphi}\right)^2 
\quad{\rm and} \quad 
V(\psi) = V_0 \sinh^2(\psi/\sqrt{2p}) \, ,
\label{pot1}
\end{equation}
where 
\begin{equation}
V_0 = p\,m^2m_P^2 \quad{\rm and}\quad V_C = \frac{3}{4}\alpha m_\phi^2m_P^2,
\label{pot2}
\end{equation}
where $p = 3\beta$ and we have reinstated $m_P$. When $\alpha = 1$, then
$V(\varphi)$ becomes the Starobinsky potential. We emphasise here that this is 
but an example of a plateau model of slow-roll GUT-scale plateau-inflation 
$V(\varphi)$; the so-called E-model inflation \cite{Emodel}. As another example,
we may consider T-model inflation instead, where the denominator in the 
potential for $\phi$ in Eq.~(\ref{eq:alphaattractor}) is missing and 
\mbox{$V(\varphi)\propto(\tanh\frac{\varphi/m_P}{\sqrt{6\alpha}})^2$} 
\cite{Tmodel}.\footnote{
In fact, any plateau model would do (e.g. Ref.~\cite{mine}).}

As shown in Refs.~\cite{Broy:2015qna,Rinaldi:2015yoa,Terada:2016nqg} the 
$\alpha$-attractors approach 
works also for other forms of the pole {of a kinetic term} and it may be responsible for different 
forms of potentials, e.g. without inflationary plateau. This is exactly the 
case of the potential in the $\psi$ direction - instead of plateau we obtain an
exponential potential, which comes from the pole of $V(s)$. Indeed,
for $|\psi|\gg\psi_c$ (where $\psi_c\equiv\sqrt{2p}\,m_P$) one finds 
\footnote{An exponential potential for proto-inflation was also obtained in 
Ref.~\cite{Carrasco:2015rva}. However, in contrast to 
Ref.~\cite{Carrasco:2015rva}, the fields $\varphi$ and $\psi$ in our model
are unrelated and may belong to entirely different sectors of the theory.}
\begin{equation}
V(\psi)\simeq V_0\exp\left(\mbox{$\sqrt{\frac{2}{p}}$}\,|\psi|/m_P\right) \label{eq:exp}
\end{equation}
and therefore it is equivalent to the potential of power-law inflation with 
$a\propto t^p$. In the $|\psi|\ll\psi_c$ one finds
\begin{equation}
V(\psi) \simeq \frac{1}{2}m^2 \, \psi^2 \, ,
\end{equation}
which is the potential of the quadratic chaotic model, for which inflation ends
around $|\psi|\simeq \sqrt 2\,m_P$. Note that, with \mbox{$p\gsim 1$}, 
one finds $V(\pm\psi_c)\simeq V_0$. The form of the potential is shown in 
Fig.~\ref{fig}.

The model of inflation considered has 3 phases: 
\begin{itemize}
\item[1:] 
Power-law proto-inflation driven by $\psi$, which starts at the Planck scale, 
\vspace{-.3cm}
\item[2:] 
Quadratic chaotic inflation, for $m_P<|\psi|<\psi_c$, 
\vspace{-.3cm}
\item[3:] 
Inflation on the GUT-scale plateau generated by $\varphi$. 
\end{itemize}
In order to fit the model to the data, phase 3 needs to be the last phase 
of inflation and therefore it is preferable to avoid eternal inflation with 
an extended diffusion zone in the inflationary phase~1. Otherwise, e.g. if 
quadratic chaotic inflation were employed in phase~1. one
would end up with prolonged eternal inflation at sub-Planckian energy scales. 
This would mean that, most likely, by the time the proto-inflaton $\psi$ would 
manage to exit the diffusion zone, the plateau-inflaton $\varphi$ 
(responsible for phase~3) could long be in its own minimum. 

Note that eternal inflation appears naturally in the context of the multiverse, 
suggested by string theory. Within our approach we are not trying to 
secure the lack of eternal inflation. We simply consider that the diffusion 
zone for proto-inflation is not extended, but it is confined to the Planck 
scale. So, after starting at Planck scale, proto-inflation avoids the diffusion 
zone during its evolution.

The potential of the proto-inflation can be easily generalised into
\begin{equation}
\mathcal{L}\ni\frac{1}{2}m^2\frac{s^{2n}}{\left(1-\frac{s^2}{6\beta}\right)^n}\,,
\end{equation}
which, for the canonically normalised field, gives
\begin{equation}
V(\psi) = \frac{1}{2}m^2(6\beta)^n \sinh^{2n}(\psi/\sqrt{2p}) \,,
\end{equation}
with $m_P=1$.
Similarly to the original potential in Eq. (\ref{eq:alphaattractor}), 
the above (in the $\psi>\psi_c$ limit) results in an exponential potential, 
which supports power-law proto-inflation, namely
\begin{equation}
V(\psi) \propto \exp\left(n\sqrt{\frac{2}{p}}|\psi|/m_P\right) \, .
\end{equation}
After the re-definition $p\to n^2p$, one recovers Eq. (\ref{eq:exp}). In 
the $m_P\lsim |\psi|<\psi_c$ limit one obtains a $V \propto \psi^{2n}$ chaotic 
inflation. Therefore any order of the pole can be used to generate 
proto-inflation without an extended diffusion zone from the Planck scale to the
GUT scale. The importance of the pole in the proto-inflationary potential should
be stressed again. Without it, $V(\psi)$ would be too flat to avoid eternal 
inflation at sub-Planckian scales.

{Another example of obtaining non-flat potential from 
$\alpha$-attractors would be to consider 
\begin{equation}
\mathcal{L}\ni V_0 
\ln\left(2\,\frac{6\beta + s^2}{6\beta - s^2}\right)=
V_0\ln[2\cosh(2\psi/\sqrt{6\beta})]\,,
\end{equation}
where $m_P=1$ again. In such a case, 
one finds a potential with smooth minimum at $s = \psi = 0$ with two arms that 
correspond to $V(\psi)\simeq 2V_0|\psi|/\sqrt{6\beta}$, when $\psi^2\gg 6\beta$ 
(i.e. $s^2\rightarrow 6\beta$). Proto-inflation would be linear inflation, 
which however is not free of the extended eternal inflation problem. Therefore,
we mention this model only as another example of how a pole of the potential may
generate an $\alpha$-attractor model without a plateau.}

\subsection{Inflationary e-folds and the potential density scales}

The number of e-folds generated during a certain phase of inflation is 
\mbox{$N=\int_i^fHdt$},
where indices $i$ and $f$ denote the beginning and the end respectively of the 
inflationary phase in question. In power-law inflation (phase~1) we have \\
\mbox{$\psi/m_P=
\sqrt{2p}\ln\left(\sqrt{\frac{p(3p-1)}{V_0}}\frac{m_P}{t}\right)$}
\cite{DavidBook1} and \mbox{$N_1=p\ln(t_c/t_P)$}, with 
\mbox{$\psi(t_c)\equiv\psi_c$} and \mbox{$t_P=m_P^{-1}$} being the Planck 
time. In quadratic chaotic inflation we have slow-roll, so
\mbox{$N_2=m_P^{-2}\int_i^f(V/V')d\psi$}. Using the above, it is straightforward
to show that
\begin{equation}
N_{1} \simeq p \ln \frac{m_P^2}{\sqrt{V_0}}\qquad {\rm and}\qquad 
N_{2} \simeq \frac{1}{2}(p-1) \, , \label{eq:efolds}
\end{equation}
where $N_{1}$ and $N_{2}$ are the number of e-folds during phases~1 and 2 
respectively. Under the assumption \mbox{$V_0^{1/4}\sim 10^{15}\,$GeV} one finds 
$N_1\simeq 14p$. Since $p\gsim 1$, phase~2 generates no more than a handful of
e-folds.\footnote{Note that the quasi-exponential proto-inflation model in 
Ref.~\cite{Carrasco:2015rva} does not lead to any slow-roll inflation on the 
hill. }

\begin{figure}[h]
\centering

\mbox{\hspace{1cm}}

\includegraphics[width=.8\columnwidth]{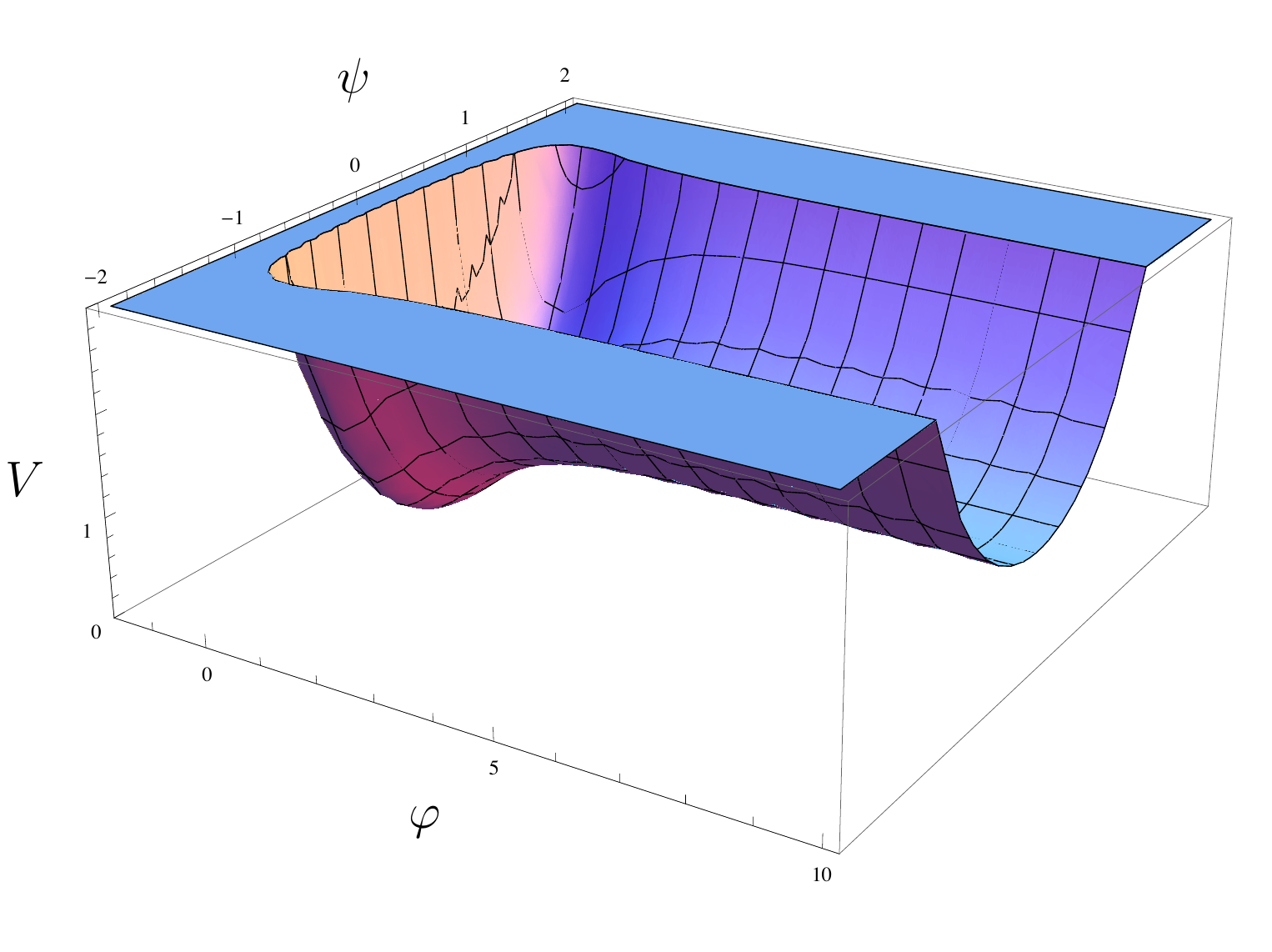}

\caption{
Schematic representation of the scalar potential 
\mbox{$V=V(\varphi)+V(\psi)$}, where $V(\varphi)$ and $V(\psi)$ are given by 
Eq.~(\ref{pot1}). First, the system undergoes proto-inflation, which is
mostly power-law (there may be some slow-roll near the end of the roll) from 
high on the slopes of the valley down to the valley itself. This is the 
$\psi$-direction. After it reaches the valley, the system undergoes GUT-scale 
slow-roll inflation towards the true minimum (on the left in the figure). This 
valley is, in fact, the inflationary plateau, which guarantees good agreement 
with Planck. As long as at least 60 e-folds of slow-roll GUT-scale inflation
along the plateau are established, the observations are satisfied. Thus, for 
the majority of the parameter space of $\varphi$ the
initial condition problem of plateau inflation is overcome, because the system 
is taken to begin with Planckian density, high on the slopes of the valley.
The valley is extremely flat for large values of $\varphi$, which means that 
one may have eternal inflation along the $\varphi$ direction. This is not a 
problem, though, because it occurs before the cosmological scales exit the 
horizon (the potential ensures that there are at least 60 e-folds of slow-roll 
GUT-scale inflation at the end; indeed it is straightforward to show that
the number of slow-roll e-folds after internal inflation in the $\phi$-direction
is \mbox{$N\simeq 3\pi\sqrt{2\alpha}(\frac{m_P}{m_{_{\rm GUT}}})^2\sim 10^6$}). 
By construction, we ensure that the $\psi$ field is heavy (i.e. $m>\frac32 H$) 
along the plateau, so no isocurvature perturbations are generated while 
$\psi$-oscillations, once the bottom of the valley is reached, are 
exponentially damped. Note that, because $\varphi$ and $\psi$ are not directly 
interacting together, $\varphi$ may be also rolling during the power-law 
inflation, but its variation cannot be much since power-law inflation lasts 
little. Assuming that $\varphi$ is not eternally inflating, any initial kinetic
energy of the plateau-inflaton $\varphi$, during proto-inflation,
is soon diluted away, so eventually $\varphi$ slow-rolls and is not 
moving much.}

\label{fig}
\end{figure}

To avoid isocurvature perturbations during phase~3, one must be sure that, 
during the $\varphi$ field domination (i.e. the GUT-scale slow-roll inflation)
$\psi$ is heavy enough not to undergo particle production. To ensure this one 
must assume
\begin{equation}
m>\frac{3}{2}H_C \  \Rightarrow \ V_0 > \frac{3p}{4}V_C \ \Rightarrow \ 
m>\frac{3}{4}\sqrt{\alpha}\,m_\phi \, , \label{eq:effsingle}
\end{equation}
where $V_C=3H_C^2m_P^2$. Eq. (\ref{eq:effsingle}) implies that 
$V(\psi)$ at the end of phase~2, satisfies
\begin{equation}
\frac{V_{\rm end}}{V_C} = \frac{4}{3\alpha}\left(\frac{m}{m_\phi}\right)^2>
\frac{3}{4} \, ,
\end{equation}
where \mbox{$V_{\rm end}\equiv V(\psi_{\rm end})\simeq(m\,m_P)^2$}.
Thus, phase~3 inflation may in principle chop off the very end of the phase~2 
inflation. However, the difference would not be bigger than $1/3$ of an e-fold 
and it can be safely ignored, so Eq. (\ref{eq:efolds}) is valid.
The outline of the scenario is described in the caption of Fig,~\ref{fig}.

\section{Discussion and Summary} \label{sec:Summary}

The Planck satellite observations favour inflation models which feature a 
plateau in the scalar potential. The energy scale of this plateau is near that 
of a Grand Unified Theory (GUT). This is much smaller that the Planck energy 
scale. As a result, chaotic initial conditions are not directly applicable. 
Indeed, taking that the kinetic, gradient and potential densities are all 
comparable at Planck energies, would mean that the horizon size at GUT-scale 
(when inflation on the plateau is to begin) would include an exponentially 
large number of inhomogeneous patches, assuming decelerated expansion until 
then. This would render inflation non-viable, since the latter requires a 
roughly homogeneous and flat horizon volume to begin.

{Recent works suggest that even inhomogeneous initial conditions may lead 
to inflation. However, they assume that, before inflation, the Universe is 
(at least predominantly) expanding. They also assume that the exponentially 
many Planck-scale Hubble volumes contained in the GUT-scale Hubble patch 
are initially non-contracting, which seems to require huge fine-tuning since the
GUT-scale Hubble patch contains $\sim 10^{12}$ Planck-scale Hubble volumes.}

{Of course, one can argue that for an infinite Universe there is a non-zero
probability of a patch of the size of the GUT-scale horizon to be initially 
expanding. In fact, an infinite Universe will have infinite such patches. But 
this is flawed logic. According to this logic, in an infinite Universe, there 
is non-zero probability of a patch corresponding to the present horizon to be 
flat and homogeneous/isotopic (plus with the correct curvature perturbation) so 
we would not need inflation at all! We would have infinite such patches as well.
However, they would still be much more unlikely to explain the current Universe 
than the patches that feature GUT-scale inflation.}

Furthermore, the assumption of the expansion in the pre-inflation era is not 
really justified, {since quantum fluctuations at Planck scale are arguably 
expected to disallow sustained expansion without inflation. Thus, starting
inflation at the Planck scale is essential. Indeed,} in the old, 
chaotic picture, it was inflation itself which generated the necessary boost 
for the Universe expansion.

One way to overcome the above problems is to consider that, immediately after 
the Planck time, inflation does take hold, but it is not the plateau-type 
inflation required by observations. Instead it is some proto-inflation, which 
accounts for the origin of the Universe expansion and ensures thereby that the 
density of the Universe decreases enough for the GUT-scale slow-roll 
plateau-inflation (the one that does satisfy the observations) to occur.
Such proto-inflation, 
however, may suffer from its own problems, namely an extended diffusion zone.

Close to the Planck energies the Hubble scale is so large that quantum 
fluctuations can overcome the classical variation of the proto-inflaton field 
$\psi$ and the plateau-inflaton field~$\varphi$. Eternal inflation is desirable
for the plateau-inflaton, because the latter random-walks away from its minimum,
increasing thereby the likelihood that there will be enough plateau-inflation 
to satisfy the Planck observations after the end of proto-inflation. In 
contrast, eternal inflation for the proto-inflaton $\psi$ can be detrimental
if it continues even at energies that are substantially sub-Planckian. The 
reason is that, with a low Hubble-scale $H$, the magnitude of the 
``quantum kicks'' of the plateau-inflaton $\varphi$ (determined by $H$) is 
diminished. As a result, the plateau-inflaton variation is more likely to 
become dominated by classical roll, which can eventually bring it to its 
minimum, even before the end of proto-inflation. Thus, it is desirable that
the diffusion zone for the proto-inflaton (i.e. the region where quantum 
fluctuations dominate the variation of $\psi$) does not extend much lower 
below $m_P$, such that the corresponding Hubble scale remains high when $\psi$ 
undergoes eternal inflation.

In other words, the problem with eternal proto-inflation at sub-Planckian scales
is that the diffusion zone for the plateau-inflaton $\varphi$ withdraws away 
from its minimum, such that there is danger that the duration of proto-inflation
may surpass the rolling period of the plateau-inflaton field $\varphi$, which 
may happen in parallel. As a result, there could be no more plateau-inflation 
left, once proto-inflation ends. 
One may address this by coupling the two inflaton fields $\psi$ and $\varphi$ 
such that before the end of proto-inflation, the plateau-inflaton is kept 
from rolling. However, such a model becomes rather restrictive because it 
requires the appropriate design.

In this paper we have presented a different solution. We have considered 
power-law proto-inflation, which does not feature an extended diffusion zone
on the hill and, once sub-Planckian, lasts only limited e-folds that serve to 
both generate the Universe expansion and also arrange safely the smooth initial
conditions for the onset of the necessary GUT-scale slow-roll plateau-inflation,
without the need to couple the two inflaton fields. 

We have realised our idea in the context of $\alpha$-attractors.
In this implementation, the two inflatons both have non-canonical kinetic terms 
featuring a pole. Such kinetic terms are well motivated in string theory and 
supergravity. The scalar potential for the associated canonically normalised 
fields is characterised by the desired shape, namely it features an inflationary
plateau (the $\varphi$-direction), which corresponds to a valley with 
exponentially steep walls that give rise to power-law proto-inflation 
(the $\psi$-direction), see Fig.~\ref{fig}. The mechanism behind the 
$\alpha$-attractors is generating the plateau by stretching the potential
in the $\varphi$ direction. In the $\psi$ direction, however, the same mechanism
does not result in a plateau because the potential of the associated
non-canonical degree of freedom features {\em the same pole} 
of any order 
as the corresponding kinetic term.

As mentioned, eternal inflation along the $\varphi$ valley is desirable because
it pushes the expectation value of the plateau-inflaton away from its minimum.
Still, many authors have considered eternal inflation as a problem because 
``everything that is possible will happen, an infinite number of times''. This,
however, is not a problem in our simple model, because the only thing that may 
happen is that the plateau-inflaton $\varphi$ exits the quantum diffusion zone 
and begins slow-rolling. Whether this happens soon after proto-inflation ends 
or after a huge number of eternal inflation e-folds on the plateau matters 
little. It also happens again and again (an infinite number of times) at 
different locations, but this is also irrelevant.\footnote{In 
Ref.~\cite{mukhanov}, the motivation for avoiding inflationary reproduction 
(that is eternal inflation) is that the multiverse hypothesis undermines
inflation predictability. Our work does not suffer from this problem because,
in our model, there is only one vacuum (in contrast to the multiverse hypothesis
which features $10^{500}$ or so, different vacua) so all the pocket universes
generated through eternal inflation are the same (and ours is one of them).
Thus, predictability is not undermined.} We have to live in a part of 
the cosmos where $\varphi$ exited the diffusion zone and slow roll inflation 
followed, along with reheating and the hot big bang. It is easy to show that 
$\sim 10^6$ e-folds of slow-roll inflation follow eternal inflation in the 
$\varphi$-direction. 

In fact, the only problem might occur when there is no eternal inflation along
the plateau but, after the end of power-law proto-inflation along the 
$\psi$-direction, the slow-roll plateau-inflation along the $\varphi$-direction
is too little (although you would need to be really unlucky to fall at a point 
that allows less than 60 slow-roll e-folds, out of the $10^6$ possible). With 
mild anthropic selection, we require about 60 e-folds of slow-roll inflation 
(for galaxies to have time to form), so the system should not reach 
the valley too close to \mbox{$\varphi_{\rm end}\sim\sqrt\alpha\,m_P$}; the value 
that ends GUT-scale slow-roll plateau-inflation. However, this requirement is 
not very restrictive because the value of $\varphi$ can be unlimited. Recall 
that it is the non-canonically normalised field $\phi$ that should not be 
largely super-Planckian. Transposing the kinetic pole to infinity allows 
$\varphi$, the canonically normalised field, to be arbitrary large; 
exponentially larger than $m_P$ if need be (cf. Eq.~(\ref{eq:alphakin})).

Another possible issue discussed in Ref.~\cite{Ijjas:2013vea} is that no other 
inflation should spoil our scenario by occurring in parallel. In that way,
inflation along simple monomial directions should not take place, since 
it is disfavoured by the data. Given the richness of fundamental theories the 
authors of Ref.~\cite{Ijjas:2013vea} claim that it is unlikely that inflation 
on the plateau (let alone power-law proto-inflation) is allowed to occur.
However, the only requirement is actually to consider that no other degree
of freedom apart from $\varphi$ is light, i.e. has a sub-Hubble mass. This is 
simply saying that the plateau valley has steep walls in all other directions
except $\varphi$.  That is easier to accept than the opposite,
especially in supergravity and string theory where you need special 
constructions to maintain flat directions (e.g. resolve the $\eta$-problem).

Our model is minimal because it follows the Occam's razor (parsimony) principle,
so fundamental in science, in that it invokes only two uncoupled scalar fields
without fine-tunings to account for inflationary initial conditions in a 
natural way.

\section*{Acknowledgements}
We would like to thank A.~Linde for illuminating discussions.
KD was supported (in part) by the Lancaster-Manchester-Sheffield 
Consortium for Fundamental Physics under STFC grant ST/L000520/1. 
MA was supported by the National Science Centre under research grant 
DEC-2014/13/N/ST2/02712.


\begin{thebibliography}{99}

\bibitem{Guth:1980zm}
  A.~H.~Guth,
  Phys.\ Rev.\ D {\bf 23} (1981) 347.

\bibitem{Starobinsky:1980te}
  A.~A.~Starobinsky,
  Phys.\ Lett.\ B {\bf 91} (1980) 99.

\bibitem{Ade:2015lrj}
  P.~A.~R.~Ade {\it et al.} [Planck Collaboration],
  arXiv:1502.02114 [astro-ph.CO].

\bibitem{Emodel}
  R.~Kallosh, A.~Linde and D.~Roest,
  JHEP {\bf 1311} (2013) 198;
  Phys.\ Rev.\ Lett.\  {\bf 112} (2014) no.1,  011303.

\bibitem{Tmodel}
  S.~Ferrara, R.~Kallosh, A.~Linde and M.~Porrati,
  Phys.\ Rev.\ D {\bf 88} (2013) no.8,  085038;
  JCAP {\bf 1311} (2013) 046.

\bibitem{Bezrukov:2007ep}
  F.~L.~Bezrukov and M.~Shaposhnikov,
  Phys.\ Lett.\ B {\bf 659} (2008) 703.

\bibitem{Ijjas:2013vea}
  A.~Ijjas, P.~J.~Steinhardt and A.~Loeb,
  Phys.\ Lett.\ B {\bf 723} (2013) 261.

\bibitem{East:2015ggf}
  W.~E.~East, M.~Kleban, A.~Linde and L.~Senatore,
  JCAP {\bf 1609} (2016) no.09,  010.
   
\bibitem{Clough:2016ymm}
  K.~Clough, E.~A.~Lim, B.~S.~DiNunno, W.~Fischler, R.~Flauger and S.~Paban,
  arXiv:1608.04408 [hep-th].
  
\bibitem{Kleban:2016sqm}
  M.~Kleban and L.~Senatore,
  JCAP {\bf 1610} (2016) no.10,  022
  doi:10.1088/1475-7516/2016/10/022
  [arXiv:1602.03520 [hep-th]].
  
\bibitem{Brout:1979bd}
  R.~Brout, F.~Englert and E.~Gunzig,
  Gen.\ Rel.\ Grav.\  {\bf 10} (1979) 1.
  doi:10.1007/BF00757018

\bibitem{Zelkin}
L.~Grischuk. 1981. Za.~B. Zeldovich, 
In *Moscow, Proceedings, Quantum Gravity*, 71-85;
A.~Vilenkin,
  Phys.\ Lett.\  {\bf 117B} (1982) 25.
  doi:10.1016/0370-2693(82)90866-8

\bibitem{mukhanov}
V.~Mukhanov,
  Fortsch.\ Phys.\  {\bf 63} (2015) 36
  doi:10.1002/prop.201400074
  [arXiv:1409.2335 [astro-ph.CO]].

\bibitem{Ambjorn:2012jv}
  J.~Ambjorn, A.~Goerlich, J.~Jurkiewicz and R.~Loll,
  Phys.\ Rept.\  {\bf 519} (2012) 127
  doi:10.1016/j.physrep.2012.03.007
  [arXiv:1203.3591 [hep-th]].

\bibitem{Guth:2013sya}
  A.~H.~Guth, D.~I.~Kaiser and Y.~Nomura,
  Phys.\ Lett.\ B {\bf 733} (2014) 112.

\bibitem{Dalianis:2015fpa}
  I.~Dalianis and F.~Farakos,
  JCAP {\bf 1507} (2015) no.07,  044.
 
\bibitem{Linde:2004nz}
  A.~D.~Linde,
  JCAP {\bf 0410} (2004) 004.

\bibitem{Gorbunov:2014ewa}
  D.~S.~Gorbunov and A.~G.~Panin,
  Phys.\ Lett.\ B {\bf 743} (2015) 79.

\bibitem{Hamada:2014wna}
  Y.~Hamada, H.~Kawai, Oda,~K.~Y. and S.~C.~Park,
  Phys.\ Rev.\ D {\bf 91} (2015) 053008.

\bibitem{Carrasco:2015rva}
  J.~J.~M.~Carrasco, R.~Kallosh and A.~Linde,
  Phys.\ Rev.\ D {\bf 92} (2015) 6,  063519.

\bibitem{Artymowski:2016ikw}
  M.~Artymowski, Z.~Lalak and M.~Lewicki,
  JCAP {\bf 1701} (2017) no.01,  011
  doi:10.1088/1475-7516/2017/01/011
 
\bibitem{Lucchin:1984yf}
  F.~Lucchin and S.~Matarrese,
  Phys.\ Rev.\ D {\bf 32} (1985) 1316.
  doi:10.1103/PhysRevD.32.1316

  
\bibitem{DavidBook1}
A.~R.~Liddle and D.~H.~Lyth,
``Cosmological inflation and large scale structure,''
Cambridge, UK: Univ. Pr. (2000) 400 p

\bibitem{eternal}
A.~D.~Linde,
  Mod.\ Phys.\ Lett.\ A {\bf 1} (1986) 81.

\bibitem{Linde:2016uec}
  A.~Linde,
  arXiv:1612.04505 [hep-th].

\bibitem{mine}
K.~Dimopoulos,
  Phys.\ Lett.\ B {\bf 735} (2014) 75;
  PoS PLANCK {\bf 2015} (2015) 037;
K.~Dimopoulos and C.~Owen,
  Phys.\ Rev.\ D {\bf 94} (2016) no.6,  063518.
  
\bibitem{Broy:2015qna}
  B.~J.~Broy, M.~Galante, D.~Roest and A.~Westphal,
  JHEP {\bf 1512} (2015) 149.
 
\bibitem{Rinaldi:2015yoa}
  M.~Rinaldi, L.~Vanzo, S.~Zerbini and G.~Venturi,
  Phys.\ Rev.\ D {\bf 93} (2016) 024040.

\bibitem{Terada:2016nqg}
  T.~Terada,
  Phys.\ Lett.\ B {\bf 760} (2016) 674.
 
\end{thebibliography}
\end{document}